\documentclass[twocolumn,showpacs,preprintnumbers]{revtex4}%
\usepackage{graphicx}
\usepackage{amsmath}
\usepackage{amsfonts}
\usepackage{amssymb}%
\providecommand{\U}[1]{\protect\rule{.1in}{.1in}}
\providecommand{\U}[1]{\protect\rule{.1in}{.1in}}
\providecommand{\U}[1]{\protect\rule{.1in}{.1in}}
\providecommand{\U}[1]{\protect\rule{.1in}{.1in}}
\begin{document}
\preprint{ }
\title{Fabrication of Nb/Al(AlO\textit{x})/Nb DC SQUID by focused ion beam sculpturing}
\author{D. Yuvaraj}
\author{Gil Bachar}
\author{Oren Suchoi}
\author{Oleg Shtempluck}
\author{Eyal Buks}
\email{eyal@ee.technion.ac.il}
\affiliation{Department of Electrical Engineering, Technion Israel Institute of Technology,
Haifa 32000, Israel}
\date{\today}

\begin{abstract}
Highly reproducible Nb/Al(AlO\textit{x})/Nb Josephson junction based direct
current superconducting quantum interference devices (DC SQUID) were
fabricated by three dimensional etching using focused ion beam. Hysteretic and
non-hysteretic DC SQUID with critical current ranging from $25$ to $1100%
\operatorname{\mu A}%
$ were fabricated by varying the Al barrier and oxygen exposure time. The
fabricated DC SQUIDs have shown periodic flux dependence with high modulation
factor reaching a value of $92\%$ at $4.2%
\operatorname{K}%
$.

\end{abstract}
\maketitle

Direct current superconducting quantum interference devices (DC SQUIDs) are
formed by enclosing two Josephson junction (JJs) in a superconducting loop.
They were primarily used as sensitive magnetic flux detectors and as voltage
standards, but in recent years demonstration of SQUIDs as nanoscale position
sensors, qubit readout and scanning SQUID microscopy has reopened the interest
in these devices \cite{Clarke2004,Etaki_785,Lee_841,Finkler_1046}. SQUIDs
based on Nb/Al(AlO\textit{x})/Nb JJ have relatively high transition
temperature ($T_{\mathrm{c}}$), high flux-voltage modulation factor and good
thermal recyclability and can be fabricated with a wide range of critical
currents ($I_{\mathrm{c}}$) \cite{Morohashi_1179}. Traditionally
Nb/Al(AlO\textit{x})/Nb SQUIDs with size ranging from few to several hundred
microns are fabricated in sequence of steps involving several photolithography
and anodization processes, and it also requires in-situ etching for the
deposition of top Nb electrode \cite{Morohashi_1179}. In spite of the superior
properties of Nb/Al(AlO\textit{x})/Nb SQUIDs, the fact that a
multistep-process is needed for device fabrication has limited its popularity.

The advancement in the nanofabrication techniques lead to the realization of
nano sized SQUIDs based on Al/AlO\textit{x}/Al junctions and Nb nanobridges
\cite{Finkler_1046,Troeman_2152}. SQUIDs based on Al/AlO\textit{x}/Al are
fabricated using a single shadow evaporation process, but the relatively low
transition temperature of Al limits the range of possible applications
\cite{Finkler_1046}. Although the SQUIDs based on the Nb nanobridges have
relatively high $T_{\mathrm{c}}$ and are relatively easy to fabricate, these
SQUIDs are highly hysteretic and the nanobridges typically lose their
sinusoidal current phase relationship when cooled to very low temperatures
\cite{Hasselbach_4432,Segev_104507}.Compared to alternative nanofabrication
techniques, focused ion beam (FIB) offers the possibility to etch the samples
in 3D \cite{Bell_630}. Recently Watanabe \textit{et al.} have demonstrated the
application of 3D FIB etching for the fabrication of JJs and single electron
transistors (SET) \cite{Matsuba_074303,Watanabe_410}.%

\begin{figure}
[ptb]
\begin{center}
\includegraphics[
height=1.8589in,
width=3.1116in
]%
{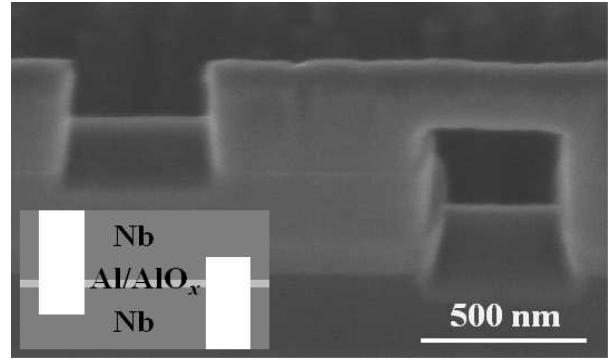}%
\caption{Scanning electron microscope (SEM) image of the Nb/Al(AlO\textit{x}%
)/Nb Josephson Junction (JJ) fabricated by 3D FIB etching. The inset shows a
schematic of JJ structure.}%
\label{device}%
\end{center}
\end{figure}

We report here a process to fabricate Nb/Al(AlO\textit{x})/Nb junction based
DC SQUIDs using only a single photolithography step and a single 3D FIB
etching. Moreover, we present the characteristics of the fabricated hysteresis
and non hysteresis DC SQUIDs at low temperature ($4.2%
\operatorname{K}%
$ and $0.1%
\operatorname{K}%
$).

Nb/Al(AlO\textit{x})/Nb trilayer was deposited on SiN coated Si substrate at
an Ar pressure of $5\times10^{-3}%
\operatorname{mbar}%
$ at room temperature by DC magnetron sputtering. Thickness of the base and
top Nb electrodes was fixed as $350%
\operatorname{nm}%
$, whereas the thickness of the Al barrier and oxidation conditions were
varied for different samples. SQUID loop, magnetic flux line and bonding pads
are patterned on the trilayer films using a single photolithography step. The
SQUID loop has an area of $40\times40%
\operatorname{\mu m}%
^{2}$ and the flux lines were fixed $10%
\operatorname{\mu m}%
$ away from the SQUID loop. JJs in the DC SQUIDs were fabricated using the
process similar to the one used by Watanabe \textit{et al.} for the
fabrication of trilayer SET \cite{Watanabe_410}. The sample with the trilayer
SQUID loop is mounted on wedge shaped ($45^{0}$) sample holder. In the 3D FIB
etching process the JJs were fabricated on the arms of the SQUID loop by
etching in perpendicular and parallel directions to the sample surface plane.
Initially using perpendicular etching with $2.8$ $%
\operatorname{nA}%
$ Ga ion current a 1 micron wide section is formed on each arm of the SQUID
loop. The width of these sections is further reduced down to $300%
\operatorname{nm}%
$ using a current of $28$ pA. Then JJ's were fabricated on the trilayer bridge
by parallel etching, in which the top and the bottom Nb layer were etched with
$1.5$ pA Ga ion current as schematically shown in the inset of Fig.
\ref{device}. Two nominally identical JJs as shown in Fig. \ref{device} with
lateral dimension of $0.12%
\operatorname{\mu m}%
^{2}$ were fabricated in each arm of the SQUID loop.%

\begin{table}[tbp] \centering
\begin{tabular}
[c]{|l|l|l|l|l|l|}\hline
Device & Nb/Al/Nb ($%
\operatorname{nm}%
$) & $\frac{D}{%
\operatorname{torr}%
\operatorname{s}%
}$ & $\frac{I_{\mathrm{c}}}{%
\operatorname{\mu A}%
}$ & $\frac{\Delta I_{c}}{%
\operatorname{\mu A}%
}$ & $\frac{J_{\mathrm{a}}}{%
\operatorname{kA}%
\operatorname{cm}%
^{-2}}$\\\hline
DS-A & $350$/$5$/$350$ & $4.8$ & $750$ (H) & $25$ & $300$\\
DS-B & $350$/$5$/$350$ & $4.8$ & $390$ (NH) & $-$ & $160$\\
DS-C & $350$/$5$/$350$ & $12$ & $334$ (NH) & $-$ & $130$\\
DS-D & $350$/$5$/$350$ & $480$ & $25$ (NH) & $-$ & $2.4$\\
DS-E & $350$/$3$/$350$ & $6$ & $1070$ (H) & $70$ & $450$\\\hline
\end{tabular}
\caption{The parameters of the different DC SQUIDs fabricated using FIB. Device DS-A after
anodization process is labeled as device DS-B. The parameter $D=P\tau$, where $P$
is the oxygen partial pressure and $\tau$ is the oxygen exposure time, characterizes
the oxidation doze. In the $I_{\mathrm{c}}$ column H and NH stands for 'hysteretic'
and 'non-hysteretic' respectively. The parameter $\Delta I_{\mathrm{c}}$
is given by $\Delta I_{\mathrm{c}}=I_{\mathrm{c}1}-I_{\mathrm{c}2}$, where
$I_{\mathrm{c}1}$ and $I_{\mathrm{c}2}$ are the critical currents measured in
the sweep up and sweep down directions respectively.
\label{key}}%
\end{table}%

I-V characteristics of the DC SQUIDs with different Al barrier thickness and
oxidation conditions were measured at $4.2%
\operatorname{K}%
$ and the parameters of these devices are tabulated in table 1. The critical
current of the SQUID decreased from $750$ to $25%
\operatorname{\mu A}%
$ with the increase in the oxygen exposure time. As the fabricated JJs are of
similar size the observed behavior is due to the decrease in the critical
current density ($J_{\mathrm{a}}$) of the junction with the increases in the
oxygen exposure time, which determines the AlO$_{x}$ barrier thickness. The
hysteresis present in the devices having relatively thin AlO$_{x}$ barrier
thickness is removed with the increases in the oxygen exposure time. The
current density of the junctions increased with the reduction in the Al
barrier thickness as observed in SQUID DS-E. In all the DC SQUIDs except DS-B
and DS-D, we have fabricated 6 identical SQUIDs and the mean value of
$I_{\mathrm{c}}$ is tabulated in the table.1. The $I_{\mathrm{c}}$\ of the DC
SQUIDs was found to be highly reproducible with less than $\pm6\%$ variation
from the mean value.%

\begin{figure}
[ptb]
\begin{center}
\includegraphics[
height=2.3653in,
width=3.1116in
]%
{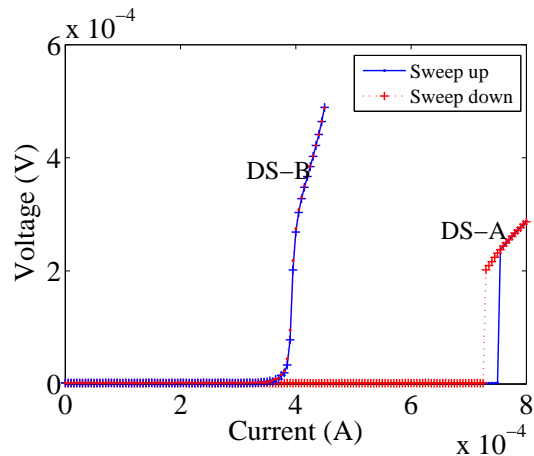}%
\caption{The I-V characteristic of the as-fabricated (DS-A) and anodized
(DS-B) Nb/Al(AlO\textit{x})/Nb DC SQUIDs measured at $4.2\operatorname{K}$.}%
\label{IV}%
\end{center}
\end{figure}

\qquad Figure \ref{IV} shows the current-voltage (I-V) characteristics of a DC
SQUID (DS-A) measured at $4.2%
\operatorname{K}%
$. The I-V curve shows that the SQUID is hysteretic with a critical current
($I_{\mathrm{c}}$) of $750%
\operatorname{\mu A}%
$. The hysteretic behavior of the SQUID indicates that the junctions are
underdamped \cite{Kivioja_179}. During the FIB sidewall etching, the Ga ions
implanted into the sidewalls suppress the superconductivity of Nb up to a
thickness of about $30%
\operatorname{nm}%
$, as estimated using simulation by Troeman \textit{et al.}
\cite{Troeman_2152}. To avoid any contribution to the $I_{\mathrm{c}}$ of the
tunnel junction from this Ga implanted sidewall, they were passivated using
standard anodization process \cite{kroger_280}. DC SQUIDs were anodized at
constant potential of $40%
\operatorname{V}%
$ for $60%
\operatorname{s}%
$ in an electrolytic solution containing $165%
\operatorname{g}%
$ ammonium pentaborate, $1120%
\operatorname{ml}%
$ ethylene glycol and $760%
\operatorname{ml}%
$ deionized water. Figure \ref{IV} shows the I-V characteristic of the DC
SQUID after anodization (DS-B), which resulted in reduction of the
$I_{\mathrm{c}}$ of the SQUID from $750$ to $390%
\operatorname{\mu A}%
$, and in the elimination of the hysteretic behavior. The normal state
resistance of the junctions determined from the I-V increased from 1 to $4.8%
\operatorname{\Omega }%
$ after anodization.%

\begin{figure}
[ptb]
\begin{center}
\includegraphics[
height=2.3761in,
width=3.1116in
]%
{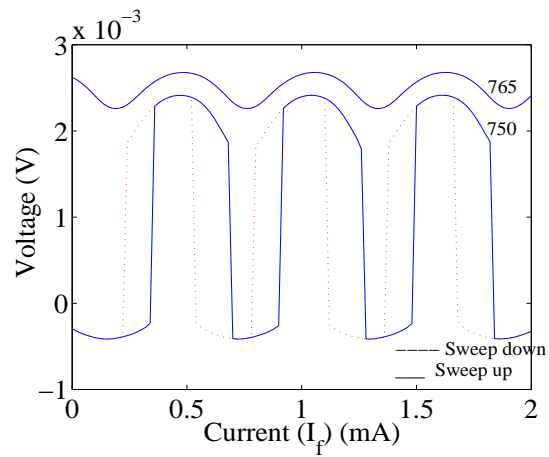}%
\caption{Flux response of the hysteretic (DS-A) DC SQUID measured for
$I_{\mathrm{b}}$ of $750$ and $765\operatorname{\mu A}$ in sweep up and sweep
down directions.}%
\label{hysteretic}%
\end{center}
\end{figure}

The flux response of the hysteretic (DS-A) and non-hysteretic (DS-B) DC SQUIDs
were measured at $4.2%
\operatorname{K}%
$. In this measurement the modulation in the voltage across the SQUID was
recorded as a function of the current applied to the magnetic flux lines
($I_{\mathrm{f}}$) and SQUID bias current ($I_{\mathrm{b}}$). The flux
response of the hysteretic SQUID was measured from $I_{\mathrm{f}}=0%
\operatorname{mA}%
$ to $2%
\operatorname{mA}%
$ in steps of $0.02%
\operatorname{mA}%
$ both in sweep up and sweep down directions. This measurement was repeated
for SQUID bias current $I_{\mathrm{b}}$ of $750$ and $765%
\operatorname{\mu A}%
$ (see Fig. \ref{hysteretic}). For $I_{\mathrm{b}}$ of $750%
\operatorname{\mu A}%
$ the voltage across the SQUID is highly modulated and measured voltage is
asymmetric function of flux both in the sweep up and down directions as shown
in Fig. \ref{hysteretic}. The voltage modulation factor of the SQUID
calculated using $((V_{\max}-V_{\min})/V_{\max})\times100\%$ in this region is
found to be $96\%$ (where $V_{\max}$ and $V_{\min}$ are respectively the
maximum and minimum voltage measured with respect to the flux changes). The
flux response curve becomes symmetric and the modulation factor decreases to
$19\%$ at $I_{\mathrm{b}}$ of $765%
\operatorname{\mu A}%
$ and the flux response curve for the sweep up and down direction become
identical as shown in Fig. \ref{hysteretic}.%

\begin{figure}
[ptb]
\begin{center}
\includegraphics[
height=2.0457in,
width=3.4006in
]%
{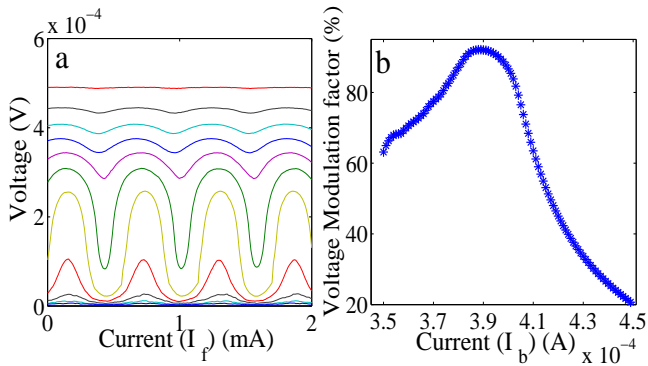}%
\caption{(a) Flux response of a non-hysteretic (DS-B) DC SQUID measured at
different SQUID bias current ($I_{\mathrm{b}}$) from $350$ to
$450\operatorname{\mu A}$ in steps of $10\operatorname{\mu A}$. (b) Plot of
voltage modulation factor as a function of SQUID bias current ($I_{\mathrm{b}%
}$).}%
\label{non-hysteretic}%
\end{center}
\end{figure}

The flux response of the non-hysteretic DC SQUID (DS-B) is shown in Fig.
\ref{non-hysteretic}a. This SQUID did not exhibit flux sensitivity for
$I_{\mathrm{b}}$ below $350$ $%
\operatorname{\mu A}%
$ and above $450%
\operatorname{\mu A}%
$, and showed periodic modulation of the voltage within this range as shown in
Fig. \ref{non-hysteretic}a. In this non-hysteretic SQUID no switching like
behavior is observed and the voltage is rather smooth and symmetric with
respect to $V_{\max}$ with relatively high modulation depth. The flux response
curves of this SQUID are identical irrespective of the direction of the
$I_{\mathrm{f}}$ sweep and hence the flux response with respect to sweep up
direction is only plotted here. The calculated modulation factor is plotted
against $I_{\mathrm{b}}$ in Fig. \ref{non-hysteretic}b. The modulation factor
increased from $60\%$ at $I_{\mathrm{b}}$ of $350%
\operatorname{\mu A}%
$ and reached a maximum value of $92\%$ at $I_{\mathrm{b}}$ of $392%
\operatorname{\mu A}%
$. After reaching the maximum the modulation factor gradually decreases with
the further increase in $I_{\mathrm{b}}$ as shown in Fig. \ref{non-hysteretic}%
b. The relatively high modulation factor indicates that the two JJs of the
SQUID have similar critical currents, demonstrating thus the good
repeatability of the FIB fabrication technique.

The operation of the DC SQUID (DS-A) was tested from $4.2%
\operatorname{K}%
$ down to $0.1%
\operatorname{K}%
$. The critical current of the junction increased from $750%
\operatorname{\mu A}%
$ at $4.2%
\operatorname{K}%
$ to $1880%
\operatorname{\mu A}%
$ at $0.1%
\operatorname{K}%
$. The flux response of this SQUID measured at $0.1%
\operatorname{K}%
$ had shown periodic voltage modulation with a modulation factor of $80\%$.
These results demonstrate that DC SQUIDs fabricated in the present process
have very high modulation factor over a wide range of temperatures compared to
the DC SQUIDs made of Nb nanobridges.

In summary, Nb/Al(AlO\textit{x})/Nb DC SQUIDs with nanoscale JJs were
successfully fabricated using relatively simple process consisting of a single
photolithography step and a single 3D FIB etching step. Using this process
hysteretic and non-hysteretic DC SQUIDs were fabricated by varying the Al
barrier thickness and oxidation conditions. Hysteretic SQUIDs exhibited
asymmetric flux response curve with high modulation factor. Non-hysteretic
SQUIDs have shown smooth symmetric flux response, and the voltage modulation
factor in this case reached a maximum value of $92\%$ near the critical
current of the SQUID.

D. Yuvaraj is supported by the Viterbi fellowships. This work is supported by
the German Israel Foundation under grant 1-2038.1114.07, the Israel Science
Foundation under grant 1380021, the Deborah Foundation, Russell Berrie
Nanotechnology Institute, the European STREP QNEMS Project, Israeli ministry
of science and MAFAT.\qquad

\bibliographystyle{apsrev}
\bibliography{acompat,Eyal_Bib}

\bigskip

\bigskip

\bigskip

\bigskip

\bigskip

\bigskip

\bigskip

\bigskip

\bigskip

\bigskip

\bigskip

\bigskip

\bigskip

\bigskip

\bigskip

\bigskip

\bigskip

\bigskip

\bigskip

\bigskip

\bigskip

\bigskip

\bigskip

\bigskip

\bigskip

\bigskip

\bigskip
\end{document}